\documentclass[12pt]{article}
\usepackage{amsmath, amsfonts, amscd}
\usepackage{amsthm}
\newcommand{\slie}{\mathcal{L}}
\newcommand{\jl}{\mathcal{A}}

\newcommand{\com}{\mathbb{C}}
\newcommand{\A}{\mathbb{A}}
\newcommand{\mapto}{\longrightarrow}

\newcommand{\N}{\mathbb{N}}
\newcommand{\Z}{\mathbb{Z}}
\newcommand{\R}{\mathbb{R}}
\newcommand{\de}{\delta}
\newcommand{\jor}{\mathcal{J}}
\newcommand{\quat}{\mathbb{H}}
\newcommand{\cayl}{\mathbb{K}}
\newcommand{\D}{\mathbb{D}}
\newcommand{\JL}{\mathfrak{J}\mathfrak{L}}

\title{\bf `Mixed' $\mathbf{\de}$-Jordan-Lie
Superalgebra}

\author{I. Raptis\thanks{Theoretical Physics
Group, Blackett Laboratory of Physics, Imperial College of
Science, Technology and Medicine, Prince Consort Road, South
Kensington, London SW7 2BZ, UK; e-mail: i.raptis@ic.ac.uk}}

\date{}

\begin{document}

\maketitle

\begin{abstract}

\noindent An algebra $\A$ not encountered in either the usual
algebraic varieties or supervarieties is introduced. $\A$ is a
$\Z_{2}$-graded and multiplicatively deformed version of the
quaternions, with structure similar to that of a $\de$-Jordan-Lie
algebra as defined in \cite{okam}, but it is shown to be neither
that of a purely associative ($\de=+1$) Lie superalgebra, nor that
of a purely antiassociative ($\de=-1$) Jordan-Lie superalgebra.
Rather, it exhibits a novel kind of associativity, here called
`{\em ordered $\Z_{2}$-graded associativity}', that is somewhat
`in between' pure associativity and pure antiassociativity. In
addition to graded associativity, the generators of $\A$ obey
graded commutation relations encountered in both the usual
$\Z_{2}$-graded Lie algebras ($\de=1$) and in $\Z_{2}$-graded
Jordan-Lie algebras ($\de=-1$). They also satisfy new graded
Jacobi identities that combine characteristics of the Jacobi
relations obeyed by the generators of ungraded Lie,
$\Z_{2}$-graded Lie and $\Z_{2}$-graded Jordan-Lie algebras.
Mainly due to these three features, $\A$ is called a `{\em mixed'
$\de$-Jordan-Lie superalgebra}. The present paper defines $\A$ and
compares it with the $\de$-Jordan-Lie algebra defined in
\cite{okam}.

\end{abstract}

\newpage

\section{Introduction}\label{sec1}

In theoretical physics, supersymmetry pertains to a symmetry
between bosons and fermions. Supergroups, or $\Z_{2}$-graded Lie
groups, are the mathematical structures modelling continuous
supersymmetry transformations between bosons and fermions. As Lie
algebras consist of generators of Lie groups---the infinitesimal
Lie group elements tangent to the identity, so $\Z_{2}$-graded Lie
algebras, otherwise known as Lie superalgebras, consist of
generators of (or infinitesimal) supersymmetry transformations
\cite{freund}.

Like their ungraded Lie ancestors $L$, Lie superalgebras $\slie$

\begin{itemize}

\item (i) Are complex vector spaces that are $\Z_{2}$-graded\footnote{It is tacitly assumed
that both $\slie^{0}$ and $\slie^{1}$ in (\ref{eq1}) are linear
subspaces of $\slie$ whose only common element is the zero vector
$0$. $\slie^{0}$ is usually referred to as {\em the even subspace
of} $\slie$, while $\slie^{1}$ as {\em the odd subspace of}
$\slie$.}

\begin{equation}\label{eq1}
\slie=\slie^{0}\oplus\slie^{1},
\end{equation}

\noindent with grading function $\pi$ given by

\begin{equation}\label{eq2}
\pi(x):=\left\lbrace\begin{array}{rcl} 0,& \mbox{when}~
x\in\slie^{0},\cr 1, & \mbox{when}~ x\in\slie^{1}.
\end{array} \right.
\end{equation}

\item (ii) Are associative algebras with respect to a bilinear product
$\cdot:~\slie\otimes\slie\mapto\slie$ (simply write $x\cdot
y\equiv xy=z\in\slie$ for the associative product $\cdot$ of $x$
and $y$ in $\slie$).

\item (iii) Close under the so-called super-Lie bracket $<.
,.>:~\slie\otimes\slie\mapto\slie $ represented by the
non-associative, bilinear, $\Z_{2}$-graded (anti-commutator) Lie
product $[.,.\}$ defined as

\begin{equation}\label{eq3}
[x,y\}:=\left\lbrace\begin{array}{rcl} [x,y]=xy-yx\in\slie^{0},&
\mbox{when}~ x,y\in\slie^{0},\cr \{ x,y\}=xy+yx\in\slie^{0}, &
\mbox{when}~ x,y\in\slie^{1},\cr [x,y]=xy-yx\in\slie^{1}, &
\mbox{when}~ x\in\slie^{0}~{\rm and}~ y\in\slie^{1}.
\end{array} \right.
\end{equation}

\item (iv) With respect to $<.,.>$, they obey the so-called
super-Jacobi identities\footnote{For more details about the
properties (i)--(iv) of Lie superalgebras, the reader is referred
to \cite{freund}. We will encounter them in a slightly different
guise and in more detail when we define $\de$-Jordan-Lie
superalgebras in the next section.}.

\end{itemize}

In what follows we first recall the definition of an abstract
$\de$-Jordan-Lie ($\de$-J-L) algebra $\jl$ given by Okubo and
Kamiya in \cite{okam}, which, as we shall see, includes as a
particular case the Lie superalgebra defined in (i)-(iv) above
(section \ref{sec2}), then we introduce the concrete algebra $\A$
(section \ref{sec3}), and finally we compare the key defining
properties of the two structures (section \ref{sec4}). Section
\ref{sec5} concludes the paper with some brief remarks about a
possible physical application and interpretation of $\A$.

\section{$\de$-Jordan-Lie Superalgebra}\label{sec2}

Let $\jl$ be a finite dimensional vector space over a field $K$ of
characteristic not $2$ which, for familiarity, one may wish to
identify with $\com$. Also, let $\jl$ be $\Z_{2}$-graded

\begin{equation}\label{eq4}
\jl=\jl^{0}\oplus\jl^{1},
\end{equation}

\noindent with grader $\pi$ given by

\begin{equation}\label{eq5}
\pi(x):=\left\lbrace\begin{array}{rcl} 0,& \mbox{when}~
x\in\jl^{0},\cr 1, & \mbox{when}~ x\in\jl^{1},
\end{array} \right.
\end{equation}

\noindent as in (\ref{eq1}) and (\ref{eq2}) for $\slie$
above\footnote{In \cite{okam}, $\sigma(x)$ is used instead of
$\pi(x)$ to symbolize the grading function. See $(1.2)$ in
\cite{okam}. }.

Next, we consider only homogeneous elements of $\jl$ ({\it ie},
either $x\in\jl^{0}$ or $x\in\jl^{1}$, but not $z=\alpha x+\beta
y,~ x\in\jl^{0},~ y\in\jl^{1};~\alpha ,\beta\in \com$)\footnote{In
theoretical physics, this forbidding of linear combinations
between bosons and fermions is known as the Wick-Wightman-Wigner
superselection rule \cite{www}. The direct sum split between the
even and the odd subspaces in (\ref{eq1}) and (\ref{eq4}) is
supposed to depict precisely this constraint to free
superpositions between quanta of integer and half-integer spin
({\it ie}, bosons and fermions, respectively). Mainly because of
\cite{www} we decided to symbolize the grading function in
(\ref{eq2}) and (\ref{eq5}) by `$\pi$' (for `{\em intrinsic
parity}') rather than by `$\sigma$' as in \cite{okam}. In the
literature, the set-theoretic (disjoint) union `$\cup$' is
sometimes used instead of `$\oplus$' between the even and odd
subspaces of a $\Z_{2}$-graded vector space \cite{freund}---it
being understood that these two subspaces have only the trivial
zero vector in common, as noted in footnote 1. `$\cup$' too is
supposed to represent the aforesaid spin-statistics superselection
rule.}, and as in $(1.3)$ of \cite{okam} we define

\[
(-1)^{xy}:=(-1)^{\pi(x)\pi(y)}.
\]

Let also $xy$ be a bilinear product in $\jl$ satisfying

\begin{equation}\label{eq6}
(xy)z=\de x(yz),~\de=\pm1,
\end{equation}

\noindent with respect to which $\jl$ is said to be a {\em
$\de$-associative algebra}. In particular, for $\de=+1$, $\jl$ is
an {\em associative} algebra; while for $\de=-1$, it is {\em
antiassociative}.

Consider also a second bilinear product $<.,.>:\, \jl\otimes
\jl\mapto\jl$

\begin{equation}\label{eq7}
<x,y>:=xy-\de(-1)^{xy}yx,
\end{equation}

\noindent satisfying

\begin{equation}\label{eq8}
\pi(<x,y>)=\pi(x)+\pi(y)~~({\rm mod}~ 2),
\end{equation}

\begin{equation}\label{eq9}
<x,y>=-\de(-1)^{xy}<y,x>,
\end{equation}

\noindent and

\begin{equation}\label{eq10}
(-1)^{xz}<<x,y>,z>+(-1)^{yx}<<y,z>,x>+(-1)^{zy}<<z,x>,y>=0,
\end{equation}

\noindent or equivalently

\begin{equation}\label{eq11}
(-1)^{xz}<x,<y,z>>+(-1)^{yx}<y,<z,x>>+(-1)^{zy}<z,<x,y>>=0.
\end{equation}

$\jl$, satisfying (\ref{eq4})--(\ref{eq11}), is called a {\em
$\de$-J-L algebra} \cite{okam}. Also, one can easily verify that
for $\de=1$, $\jl$ is the associative $\Z_{2}$-graded Lie
superalgebra $\slie$ defined in (i)--(iv) of section
\ref{sec1}\footnote{In particular, the expression (\ref{eq3}) in
(iii) is encoded in (\ref{eq7})--(\ref{eq9}) above, while the
`graded Jacobi identities' property (iv) of $\slie$ is expressed
by (\ref{eq10}) or (\ref{eq11}).}. The antiassociative ($\de=-1$)
case is coined {\em Jordan-Lie superalgebra} in \cite{okam}---here
to be referred to as {\em J-L algebra} $\jor$ for short. We may
summarize all this as follows

\[
\jl=\left\lbrace\begin{array}{rcl} \slie, & \mbox{for}~ \de=+1,\cr
\jor, & \mbox{for}~ \de=-1.
\end{array} \right.
\]

For future use we quote, without proof, the following lemma and
two corollaries from \cite{okam}\footnote{Proofs can be read
directly from \cite{okam}.}:

\begin{itemize}

\item {\bf Lemma:} {\em In every antiassociative algebra $A$, any product involving
four or more elements of $A$ is identically zero}\footnote{Lemma
$1.1$ in \cite{okam}.}.

\item {\bf Corollary 1:} {\em Antiassociative algebras have no
idempotent elements and, as a result, no units ({\it ie}, identity
elements)}\footnote{Corollary $1.2$ in \cite{okam}.}.

\item {\bf Corollary 2:} {\em Let $\jor$ be a J-L algebra as defined
above. Then $\jor$ is nilpotent of length at most 3} (write:
$\jor_{4}=0$)\footnote{Corollary $1.1$ in \cite{okam}.}.

\end{itemize}

\section{Introducing $\mathbf{\A}$}\label{sec3}

Let $\A$ be a $4$-dimensional vector space over $\R$ spanned by
$\mathbf{g}=\{ a,b,c,d\}$\footnote{The alphabetical symbolism of
the four basis vectors (generators) in $\mathbf{g}$ will be
explained subsequently.} and also be $2\oplus2$-dimensionally
$\Z_{2}$-graded thus

\begin{equation}\label{eq12}
\A=\A^{0}\oplus\A^{1}=\mathrm{span}_{\R}\{ a
,b\}\oplus\mathrm{span}_{\R}\{ c ,d\} .
\end{equation}

Let $\circ:~\A\otimes\A\mapto\A$ be a bilinear product in $\A$
which, in terms of $\A$'s generators in $\mathbf{g}$, is encoded
in the following multiplication table

\begin{equation}\label{eq13}
\begin{tabular}{|c||c|c|c|c|}
\hline
$\circ$ &$a$ &$b$ &$c$ &$d$ \\ \hline\hline $a$ &$a$ &$b$ &$-d$ &$-c$ \\
\hline $b$ &$b$ &$-a$ &$-d$ &$c$ \\ \hline $c$ &$c$
&$d$ &$a$ &$-b$ \\ \hline $d$ &$d$ &$-c$ &$b$ &$-a$ \\
\hline
\end{tabular}
\end{equation}

\vskip 0.05in

From table (\ref{eq13}), one can easily extract the following
information:

\begin{itemize}

\item {\em $\circ$ is not commutative}. In particular, $a$ commutes only with
$b$; while, $b$, $c$ and $d$ mutually anticommute. Moreover, $a$
is a right-identity, but not a left one.

\item {\em $\circ$ is not (anti)associative}. For example, one can evaluate

\[
c=-ad=a(bc)\not=\left\lbrace\begin{array}{rcl} (ab)c=bc=-d, &
(\de=+1);\cr -(ab)c=-bc=d, & (\de=-1).
\end{array} \right.
\]

\item $a$ and $c$ are $\sqrt{a}$, while $b$ and $d$ are
$\sqrt{-a}$.

\item The even subspace of $\A$ in (\ref{eq12}), $\A^{0}:=\mathrm{span}_{\R}\{
a,b\}$, is isomorphic to the complex numbers $\com$ if we make the
following correspondence between the unit vectors of $\A$ and
$\com$

\[
a\mapto 1,~~b\mapto i~ (i^{2}=-1).
\]

\noindent $\A^{0}$ is the subalgebra of even elements of $\A$.

\item The product of an even and an odd generator is odd, while
the product of two odd generators is even. Together with the
second observation above, we may summarize this to the following

\[
\pi(xy)=\pi(x)+\pi(y)~~({\rm mod 2}).
\]

\item The inhomogeneous vector $\mathbf{n_{1}}=b+c$ and the odd vector
$\mathbf{n_{2}}=c+d$ are nilpotent\footnote{$\mathbf{n_{1}}$
violates the aforementioned Wick-Wightman-Wigner superselection
rule \cite{www}.}.

\end{itemize}

Let us gain more insight into the non-associativity of $\circ$ by
making a formal correspondence between the `units' of $\A$ in
$\mathbf{g}$ and the standard unit quaternions $\mathbf{u}=\{ 1,
i, j, k\}$ in $\quat$

\begin{equation}\label{eq14}
\begin{array}{c}
a\mapto 1,~~ b\mapto i,\cr c\mapto j,~~d\mapto k.
\end{array}
\end{equation}

\noindent Then, one may wish to recall that the {\em associative
division algebra} $\quat$\footnote{We may write $\bullet$ for the
associative product of quaternions ({\it ie},
$\bullet:~\quat\otimes\quat\mapto \quat$), but omit it in actual
products, that is to say, we simply write $xy$ ($x,y\in\quat$). We
assumed the same thing for $x\cdot y$ in $\slie$ and $\jl$, as
well as for $x\circ y$ in $\A$ (for instance, see (ii) in section
\ref{sec1}).} can be obtained from $\com$ by adjoining
$j=\sqrt{-1}$ to the generators $\{ 1, i\}$ of $\com$ and by
assuming that it commutes with $1$: $1j=j1=j$, but it anticommutes
with $i$: $ij=-ji=k$\footnote{In fact, one assumes that by
transposing $i$ with $j$, $i$ gets conjugated:
$ij=ji^{*}=-ji\Leftrightarrow\{ i,j\}=0$ \cite{kauff}.}. Also, by
{\em assuming associativity}, one verifies that $k$ too is a
$\sqrt{-1}$ that anticommutes with both $i$ and $j$

\[
\begin{array}{c}
k^{2}=(ij)(ij)=i(ji)j=-i^{2}j^{2}=-1,\cr
ki=(ij)i=i(ji)=-i(ij)=-ik,
\end{array}
\]

\noindent thus one completes the following well-known
multiplication table for the unit quaternions

\begin{equation}\label{eq15}
\begin{tabular}{|c||c|c|c|c|}
\hline
$\bullet$ &$1$ &$i$ &$j$ &$k$ \\ \hline\hline $1$ &$1$ &$i$ &$j$ &$k$ \\
\hline $i$ &$i$ &$-1$ &$k$ &$-j$ \\ \hline $j$ &$j$
&$-k$ &$-1$ &$i$ \\ \hline $k$ &$k$ &$j$ &$-i$ &$-1$ \\
\hline
\end{tabular}
\end{equation}

If we were to emulate the extension of $\com$ to $\quat$ in the
case of $\A$, thus adjoin $c$ to $b$ in $\A^{0}\simeq\com$ and
require according to (\ref{eq13}) that they anticommute, as well
as that {\em $\circ$ is associative}, we would get

\begin{equation}\label{eq16}
d^{2}=(cb)(cb)=cbcb=-c^{2}b^{2}=-(a)(-a)=a
\end{equation}

\noindent which disagrees with entry $(4,4)$ in table
(\ref{eq13}). Similarly for $c$\footnote{Reader, try to calculate
$c^{2}=(bd)(bd)$ in a manner similar to (\ref{eq16}) above.}.
Clearly then, as also noted above, (the product $\circ$ in) $\A$
is neither associative\footnote{$\de=1$ in (\ref{eq6}).} nor
antiassociative\footnote{$\de=-1$ in (\ref{eq6}).}.

\begin{quotation}

\noindent {\bf Question:} {\em How can we obtain agreement between
products like the one in} (\ref{eq16})---which arise rather
naturally upon trying to extend $\com$ to $\A$ in the same manner
that $\com$ is extended to $\quat$---{\em with the entries of the
multiplication table} (\ref{eq13})? Evidently, we need a new
(anti)associativity-like law for $\circ$.

\end{quotation}

\noindent To this end we first define:

\vskip 0.1in

{\bf Definition 1:} A product string $w$ of generators of $\A$ in
$\mathbf{g}$ of length $l$ greater than or equal to
$3$\footnote{In {\em free algebra} jargon, such product strings
$w$ are called {\em words} and their factors, which are elements
of $\mathbf{g}$, are called {\em letters} (which, in turn, makes
$\mathbf{g}$ $\A$'s $4$-letter alphabet!). The number $l$ of
letters in a word $w$ is its {\em length}, and we write $l(w)$.
Formally speaking, a word $w$ of length $l$ is a member of
$\stackrel{l~{\rm factors}~\A}
{\overbrace{\A\otimes\A\otimes\cdots\otimes\A}}$. The $4^{2}$
possible words of length $2$ in $\A$ are the ones depicted in
table (\ref{eq13}) above.} is said to be {\em ({\sl n})ormally
({\sl o})rdered}\footnote{Write `{\sl no}-ed' and symbolize the
word by $\overleftarrow{w}$.} if it is of the following `{\em
right-to-left alphabetical order}' or `{\em alphabetic-syntax}'

\begin{equation}\label{eq17}
\overleftarrow{w}:=d^{s}c^{r}b^{q}a^{p},~~p,q,r,s\in\N;~~l(w):=p+q+r+s.
\end{equation}

\noindent Then we impose the following three {\em rules} or {\em
relations}\footnote{Again, this is free algebra jargon.} onto the
total contraction of any word of length $l\geq3$\footnote{By `{\em
total contraction}' of a word of length $l\geq3$ we mean the
reduction of the word to a single (signed) letter in $\mathbf{g}$
after $l-1$ pairwise contractions of its constituent letters
according to (\ref{eq13}). Again, formally speaking, the product
$\circ:~\A\otimes\A\mapto\A$ in (\ref{eq13}) represents the
contraction of $2$-words in $\A$, so analogously, the total
contraction of words of length $l$ may be cast as $\circ^{l-1}:~
\stackrel{l-1~{\rm times}}
{\overbrace{\A\otimes\A\otimes\cdots\otimes\A}}\mapto\A$.}:

\vskip 0.1in

{\bf Rule 0:} Before contracting totally a word $w$ of length
$l\geq3$ it should be brought into {\sl no}-ed form in the
following two steps:

\begin{itemize}

\item (a) When the right-identity letter $a$ is found in an extreme
left or intermediate position in $w$, it should be contracted with
the adjacent letter on its right according to
(\ref{eq13})\footnote{As it were, the `natural' position of $a$ in
a word is to the extreme right. This seems to suit $a$'s role as a
right-identity in $\A$ (\ref{eq13}).}.

\item (b) The other three mutually anticommuting generators $b$, $c$ and $d$
in $\mathbf{g}$ should be pairwise swapped within $w$ so that they
are ultimately brought to the form $\pm d^{s}c^{r}b^{q}$.

\end{itemize}

\noindent A couple of comments are due here:

\vskip 0.1in

1) Above, (a) implies that the length of a word may change upon
{\sl no}-ing it. This is allowed to happen in $\A$. For the
algebraic structure of $\A$ that we wish to explore here not all
words assembled by free (arbitrary) $\circ$-concatenations of
letters in $\mathbf{g}$ are significant. {\em Only {\sl no}-ed
words are structurally significant}\footnote{This will be amply
justified in the sequel.} , and any given $w$ has a unique {\sl
no}-ed form $\overleftarrow{w}$ fixed according to (i) and (ii)
above. Rule 0 prompts us to call $\A$ `{\em multiplicatively
ordered}' and this alphabetico-syntactic ordering may be formally
cast as follows

\begin{equation}\label{eq18}
d>c>b>a,
\end{equation}

\noindent since, once again, every {\sl no} word is of the form
$\overleftarrow{w}:=d^{s}c^{r}b^{q}a{p}$ according to
(\ref{eq17}). The generators of $\A$ are ordered thus. Since in
its transition to its unique {\sl no}-ed form a word may change
length, the latter is not a significant structural trait of $\A$,
but the order (\ref{eq18}) is.

2) Normal ordering respects superpositions of words in $\A$. In
other words, {\sl no}-ing is a linear operation; symbolically

\[
\overleftarrow{w_{1}+w_{2}}=\overleftarrow{w_{1}}+\overleftarrow{w_{2}}.
\]

\noindent The other two rules that we impose on the total
contraction of a {\em no}-ed word of length $l\geq3$ are:

\vskip 0.1in

{\bf Rule 1:} Every {\sl no}-ed word of length $l$ greater than 2
contracts totally to a (signed) letter in $\mathbf{g}$ by $l-1$
sequential pair-contractions of letters in it according to
(\ref{eq13}) {\em (f)rom (r)ight (t)o (l)eft}\footnote{Write
`frtl'.} ({\it ie}, in the multiplicative order depicted in
(\ref{eq18})). We may call this rule for $\circ$ {\em ordered} or
{\em directed associativity}.

\vskip 0.1in

{\bf Rule 2:} Moreover, ordered associativity is $\Z_{2}$-graded
as follows

\small{
\begin{equation}\label{eq19}
\begin{array}{l}
\overleftarrow{w_{1}}=\ldots oe^{'}e\mapto(\overleftarrow{w_{1}})=
(-1)^{[\pi(e)+\pi(e^{'})]}\ldots o(e^{'}e)=+\ldots oe^{''},~
e^{''}=(e^{'}e)~{\rm from}~(13)\cr \overleftarrow{w_{2}}=\ldots
o^{''}o^{'}o\mapto(\overleftarrow{w_{2}})=(-1)^{[\pi(o)+\pi(o^{'})]}\ldots
o^{''}(o^{'}o)=+\ldots o^{''}e,~e=(o^{'}o)~{\rm from}~(13)\cr
\overleftarrow{w_{3}}=\ldots
o^{'}oe\mapto(\overleftarrow{w_{3}})=(-1)^{[\pi(e)+\pi(o)]}\ldots
o^{'}(oe)=-\ldots o^{'}o^{''}, o^{''}=(oe)~{\rm from}~(13),
\end{array}
\end{equation}}

\noindent where $(\overleftarrow{w})$ signifies the commencement
of the pairwise sequential total contraction of the {\sl no}-ed
word $w$ frtl {\it \`a la} rule 1; `$e$' stands for ($e$)ven and
`$o$' for ($o$)dd letters in $\overleftarrow{w}$; and
`$e^{''}=(e^{'}e)~{\rm from}~(13)$' at end of the first row of
(\ref{eq19}) signifies the contraction and substitution of the
product pair $e^{'}e$ by $e^{''}$ according to
(\ref{eq13})\footnote{And from now on, $(xy)$ in $\A$ will
indicate precisely this `contraction of $xy$ and its substitution
by the corresponding entry from (\ref{eq13})' process.}. Thus,
rule 2 essentially says that {\em when an odd and an even letter
contract within a {\sl no}-ed word $\overleftarrow{w}$, one must
put a minus sign in front of $\overleftarrow{w}$}. In view of
rules 0--2, we call $\circ$ in $\A$ a `{\em $\Z_{2}$-graded
ordered associative product}'. The $\Z_{2}$-graded ordered
associativity of $\A$ is somewhat `in between' the pure
associativity of a Lie superalgebra $\slie$ ($\de=1$) and the pure
antiassociativity of a J-L algebra $\jor$ ($\de=-1$) as defined
above.

\begin{quotation}

\noindent Due to rules 0--2, $\A$ may be called a {\em
multiplicatively ordered $\Z_{2}$-graded associative
algebra}\footnote{From now on we will most often drop the adverb
`multiplicatively' above and simply refer to $\A$ as an ordered
$\Z_{2}$-graded associative algebra.}.

\end{quotation}

Having rules 0--2 in hand, we are now in a position to show that
words such as the one displayed in (\ref{eq16}) contract
consistently with the binary multiplication table (\ref{eq13}),
thus we provide an answer to the question following (\ref{eq16})
above. So, we check that

\begin{equation}\label{eq20}
d^{2}=(cb)(cb)=cbcb\stackrel{R0}{=}-c^{2}b^{2}\stackrel{R1}{=}-c^{2}(b^{2})
\stackrel{(\ref{eq13})}
{=}cca\stackrel{R1}{=}c(ca)\stackrel{R2}{=}-c^{2}\stackrel{(\ref{eq13})}{=}-a,
\end{equation}

\noindent is in agreement with (\ref{eq13})\footnote{We note that
in (\ref{eq20}), $R0$, for instance, refers to `Rule 0' (similarly
for $R1$ and $R2$). Again, for `practice' the reader can also
verify that $c^{2}=(bd)(bd)=\cdots=a$, in agreement with
(\ref{eq13}).}.

Now we can give the rest of the $\Z_{2}$-graded Lie algebra-like
structural properties of $\A$.

\begin{itemize}

\item First, there is a bilinear product
$<.,.>:~\A\otimes\A\mapto\A$ represented by the non-associative,
$\Z_{2}$-graded (anti-commutator) Lie product $[.,.\}$ as follows

\begin{equation}\label{eq21}
[x,y\}:=\left\lbrace\begin{array}{rcl} [x,y]=xy-yx\in\A^{0},&
\mbox{when}~ x,y\in\A^{0},\cr \{ x,y\}=xy+yx\in\A^{0}, &
\mbox{when}~ x,y\in\A^{1},\cr \{ x,y\}=xy+yx\in\A^{1}, &
\mbox{when}~ x\in\A^{0}~{\rm and}~ y\in\A^{1},
\end{array} \right.
\end{equation}

\noindent similar to (\ref{eq3}), or equivalently that $<.,.>$
satisfies

\begin{equation}\label{eq22}
\pi(<x,y>)=\pi(x)+\pi(y)~~({\rm mod}~2)
\end{equation}

\noindent and

\begin{equation}\label{eq23}
\begin{array}{c}
<x,y>:=xy-\de(-1)^{xy}yx=-\de(-1)^{xy}<y,x>=\cr=\left\lbrace\begin{array}{rcl}
xy-(-1)^{xy}yx,& \mbox{when}~ x,y\in\A^{0}~{\rm
or}~x,y\in\A^{1};~~\underline{\de=1},\cr xy+(-1)^{xy}yx, &
\mbox{when}~ x\in\A^{0}~{\rm and}~
y\in\A^{1};~~\underline{\de=-1},
\end{array} \right.
\end{array}
\end{equation}

\noindent similar to (\ref{eq7}), (\ref{eq8}) and
(\ref{eq9})\footnote{We will comment further on (\ref{eq21}) and
(\ref{eq22})-(\ref{eq23}) in the next section when we compare $\A$
and the $\de$-J-L algebra $\jl$ of \cite{okam}.}.

\item Second, the following eight possible super-Jacobi identities

\begin{equation}\label{eq24}
\begin{array}{rcl}
&[\{ d ,c\} ,a]+\{\{ c ,a\} ,d\}+\{\{ a,d\} ,c\}=0,\cr &[\{d,c\}
,b ]+\{\{ c,b\} ,d\}+\{\{ b,d\} ,c\}=0,\cr &\{ [a,b],d\}+\{\{
b,d\} ,a\}+\{\{ d,a\} ,b\}=0,\cr &\{ [a,b],c\}+\{\{ b ,c\}
,a\}+\{\{ c,a\} ,b\}=0
\end{array}
\end{equation}

\noindent and

\begin{equation}\label{eq25}
\begin{array}{rcl}
&\{ d,\{ c,a\}\}+\{ c,\{ a,d\}\}+[ a,\{ d , c\} ]=0,\cr &\{ d,\{ c
,b\}\}+\{ c,\{ b ,d\}\}+[b,\{ d,c\} ]=0,\cr &\{ a,\{ b,d\}\}+\{
b,\{ d,a\}\}+\{ d ,[a,b]\}=0,\cr &\{ a,\{ b ,c\}\}+\{ b,\{ c
,a\}\}+\{ c,[a,b]\}=0,
\end{array}
\end{equation}

\noindent are satisfied. These are the analogues in $\A$ of
expressions (\ref{eq10}) and (\ref{eq11}) for the $\de$-J-L
algebra $\jl$ in \cite{okam}.

\end{itemize}

In view of the novel and quite peculiar ordered $\Z_{2}$-graded
associative multiplication structure $\circ$ of $\A$ (rules 0--2),
we must specify to the reader who wishes to verify patiently that
the graded Jacobi relations (\ref{eq24}) and (\ref{eq25}) hold how
to actually contract them. To this end we define:

{\bf Definition 2:} The contraction of a super-Jacobi relation is
said to be performed `{\em ($f$)rom ($i$)nside ($t$)o
($o$)utside}'\footnote{Write $fito$.} when the inner
$<.,.>$-brackets are opened and contracted first, and then the
outer ones. Analogously, the contraction of a super-Jacobi
relation is said to be $foti$ ({\it ie}, `{\em ($f$)rom
($o$)utside ($t$)o ($i$)nside}') when the outer brackets are
opened first, then the inner ones, and then the resulting
superpositions of words of length $3$ are totally contracted
according to rules 0--2.

{\bf Scholium:} The conscientious reader can check, by using
(\ref{eq13}), that the super-Jacobi relations (\ref{eq24}) and
(\ref{eq25}) are satisfied by the $fito$ mode of contraction, but
not by the $foti$ one. For instance, also to give an analytical
example of the two kinds of contraction, we evaluate the third
expression in (\ref{eq25}) by both $fito$ and $foti$ means

\begin{equation}\label{eq26}
\begin{array}{c}
\underline{fito}:~~\{ a,\{ b,d\}\}+\{ b,\{ d,a\}\}+\{ d
,[a,b]\}=\{ a, (bd)+(db)\}+\cr\{ b, (da)+(ad)\}+\{
d,(ab)-(ba)\}\stackrel{(\ref{eq13})}{=}\{ b, d-c\}=\{ b,d\}-\{
b,c\}=0\cr \underline{foti}:~~\{ a,\{ b,d\}\}+\{ b,\{ d,a\}\}+\{ d
,[a,b]\}=a\{ b,d\}+\{ b,d\} a +b\{ d,a\}+\cr\{ d,a\}b+
d[a,b]+[a,b]d= abd+adb+bda+dba+\cr
bda+bad+dab+adb+dab-dba+abd-bad= 2(ab)d+2(da)b+\cr 2(ad)b+2bda=
2(-db+da-cb-dba)=2(c+d-d+c)=2c\not=0.
\end{array}
\end{equation}

\noindent This indicates that, by virtue of the ordered
$\Z_{2}$-graded associative product structure of $\A$,

\begin{quotation}

\noindent{\em $\A$ is a Lie superalgebra-like structure with
respect to the $fito$, but not the $foti$, mode of contraction of
its graded Jacobi relations}.

\end{quotation}

\noindent This is another peculiar feature of $\A$---an immediate
consequence of its ordered $\Z_{2}$-graded associative
multiplication idiosyncracy\footnote{In the next two sections we
will discuss in more detail these `multiplication oddities' of
$\A$.}.

\section{Comparing $\jl$ with $\A$}\label{sec4}

We can now compare $\A$ with the abstract $\de$-J-L algebra $\jl$
defined in \cite{okam}. Below, we itemize this comparison:

\begin{itemize}

\item (i) As vector spaces, both $\jl$ and $\A$ are finite dimensional and
$\Z_{2}$-graded [(\ref{eq4}), (\ref{eq12})].

\item (ii) With respect to multiplication, while $\jl$ is
$\de$-associative ({\it ie}, associative $\slie$ for $\de=1$ or
antiassociative $\jor$ for $\de=-1$), $\A$ is ordered
$\Z_{2}$-graded associative---somewhat `in between' pure
associativity and pure antiassociativity [(\ref{eq6}),
(\ref{eq17}, \ref{eq18}, \ref{eq19})].

\item (iii) With respect to $\Z_{2}$-graded commutation relations $<.,.>$, $\A$
combines characteristics of both Lie superalgebras
$\slie=\jl|_{\de=1}$ and J-L algebras $\jor=\jl|_{\de=-1}$. In
particular, as [(\ref{eq7}, \ref{eq9}), (\ref{eq21})] depict:

\begin{quotation}

\noindent ({\bf a}) $\A$ is like $\slie$ with respect to the
`homogeneous' $<.,.>$-relations obeyed by even and odd
elements\footnote{That is, even elements obey antisymmetric
commutation relations, while odd elements obey symmetric
anticommutation relations. This is a concise algebraic statement
of the celebrated {\em spin-statistics connection} \cite{pauli}.}.

\end{quotation}

\noindent while:

\begin{quotation}

\noindent ({\bf b}) $\A$ is like $\jor$ with respect to the
`inhomogeneous' commutation relations between bosons and
fermions\footnote{That is, the commutation relation between an
even and an odd element of $\A$, like in $\jor$, is symmetric
({\it ie}, anticommutator).}.

\end{quotation}

\noindent moreover:

\begin{quotation}

\noindent ({\bf c}) The $\Z_{2}$-graded $<.,.>$-relations `close'
in $\A$ in exactly the same way that they close in
$\jl$\footnote{That is, in both $\jl$ and $\A$ the homogeneous
$<.,.>$-relations close in their even subspaces, while the
inhomogeneous ones in their odd subspaces.} [(\ref{eq8}),
(\ref{eq22})].

\end{quotation}

\item (iv) The generators of $\A$, unlike those in $\jl$, obey `externally
ungraded' Jacobi relations\footnote{That is, the three external
factors $(-1)^{xz}$, $(-1)^{yx}$ and $(-1)^{zy}$ present in the
Jacobi expressions (\ref{eq10}) and (\ref{eq11}) for $\jl$ are
simply missing in the corresponding ones, (\ref{eq24}) and
(\ref{eq25}), for $\A$.}. In this formal respect, $\A$ is like an
ungraded Lie algebra $L$.

\item (v) We return a bit to the comparison of the multiplication structure of the
two algebras (ii), now also in connection with the Jacobi
relations in (iv) above, and note that for the (anti)associative
$\de$-J-L superalgebras it is immaterial whether one evaluates
their super-Jacobi relations (\ref{eq10}) and (\ref{eq11}) $fito$
or $foti$, because they are `multiplicatively unordered'
structures\footnote{That is, it does not matter in what order one
contracts pairs of generators in words of length greater than $2$
in $\jl$.}. On the other hand, as we saw in (\ref{eq26}) for
example, exactly because of the ordered $\Z_{2}$-graded
associative multiplication structure of $\A$, $fito$-contracted
Jacobis are satisfied in $\A$, but $foti$ ones are not, therefore
it crucially depends on the ordered multiplication structure
$\circ$ whether $\A$ is a Lie-like algebra ($fito$) or not
($foti$). Such a dependence is absent from the multiplicative
unordered $\slie$ and $\jor$ algebras\footnote{The `multiplicative
unorderliness' of both $\jl|_{\de=1}=\slie$ and
$\jl|_{\de=-1}=\jor$ is encoded in the (anti)associativity
relation (\ref{eq6}) imposed on their products, since on the one
hand associativity simply means that the left-to-right contraction
of a $3$-letter word is the same as the right-to-left one, while
on the other, antiassociativity means essentially the same thing
under the proviso that one compensates with a minus sign for one
order of contraction relative to the other. Both associativity and
antiassociativity however, unlike the $\Z_{2}$-graded
associativity in $\A$ (\ref{eq19}), do not depend on the grade of
the letters involved in the binary contractions within words of
length greater than or equal to $3$.}.

\item (vi) Also in connection with (v) above, we note in view of
the lemma and the two corollaries concluding section \ref{sec2}
that:

\noindent ($\alpha$) Because $\A$ is not purely antiassociative,
words of length greater than or equal to $4$ in it do not vanish
identically as they do in $\jor$ for instance\footnote{See lemma
in section \ref{sec2}.}.

\noindent ($\beta$) Like the antiassociative $\jor$, $\A$ has no
idempotents and no two-sided identity. However, as we saw in the
previous section, $\A$ has a right-identity, namely,
$a$\footnote{See corollary 1 in section \ref{sec2}.}.

\noindent ($\gamma$) As a corollary of $(\alpha$), and unlike
$\jor$, $\A$ is not nilpotent of length at most $4$.

\item (vii) Finally, in connection with (iii) and (iv) above, we note that
our choice of the symmetric anticommutator relation (as in $\jor$)
instead of the antisymmetric commutation relation (as in $\slie$)
for the inhomogeneous $<.,.>$-relations in $\A$ can be justified
as follows: had we assumed $[e,o]$ instead of $\{ e,o\}$, the
$fito$ contraction of the first super-Jacobi expression in
(\ref{eq24}) would yield

\[
\begin{array}{c}
[\{ d ,c\} ,a]+\{ [c,a],d\}+\{ [a,d],c\}=\{ c+d, d\}+\{ -c-d,
c\}=\cr \{ d,d\}-\{ c,c\}=-2a-2a=-4a\not=0,
\end{array}
\]

\noindent hence the graded Jacobi identities would not have been
obeyed by the generators of $\A$ and, as a result, the latter
could not qualify as a Lie-like algebra.

\end{itemize}

\section{Closing remarks about $\A$}\label{sec5}

Our concluding remarks about $\A$ concentrate on the following
four issues:

\begin{itemize}

\item $\mathbf{(1)}$ We compare $\A$ against the other four possible Euclidean
division rings, namely, the reals ($\R$), the complexes ($\com$),
the quaternions ($\quat$) and the octonions ($\cayl$).

\item $\mathbf{(2)}$ As a particular case of $\mathbf{(1)}$, we remark about
the ordered $\Z_{2}$-graded associative $\A$ {\it versus} the
multiplicatively unordered, because associative, quaternions
$\quat$, and we briefly comment on a possible representation of
$\A$.

\item $\mathbf{(3)}$ We abstract $\A$ to a general
mixed $\de$-Jordan-Lie superalgebra $\JL$.

\item $\mathbf{(4)}$ Finally, we discuss a possible physical
application and interpretation of $\A$ as originally anticipated
in \cite{rap1}.

\end{itemize}

\vskip 0.1in

$\mathbf{(1)}$ To make the aforesaid comparison, we first recall
how abstract algebraic structure gets lost in climbing the
dimensional ladder from $\R$ to $\cayl$:

\begin{itemize}

\item Going from $\R$ of dimension $2^{0}=1$ to $\com$ of dimension $2^{1}=2$, one loses {\em
order}\footnote{Although, one gains algebraic completeness!}.

\item Going from $\com$ of dimension $2^{1}=2$ to $\quat$ of dimension $2^{2}=4$, one loses {\em commutativity}.

\item Going from $\quat$ of dimension $2^{2}=4$ to $\cayl$ of dimension $2^{3}=8$, one loses {\em
associativity}.

\item And if one wished to extend the octonions to a ring-like structure of dimension
$2^{4}=16$, which should be coined `{\em decahexanions}' $\D$,
there would be no more abstract algebraic structure to be lost
\cite{hur, kauff}.

\end{itemize}

We then note that $\A$ combines characteristics of all those four
division rings $\R$, $\com$, $\quat$ and $\cayl$, as follows:

\begin{itemize}

\item (a) $\A$ is a vector space over $\R$.

\item (b) $\A$'s even subalgebra $\A^{0}$ is isomorphic to $\com$.

\item (c) $\A$ is a $4$-dimensional vector space like $\quat$, and its $3$-subspace spanned
by the mutually anticommuting $b$, $c$ and $d$ reminds one of the
subspace of real quaternions ({\it ie}, $\quat$ over $\R$) spanned
by the three imaginary ({\it ie}, $\sqrt{-1}$) quaternion units
$i$, $j$ and $k$\footnote{With the important difference that $c$
in $\A$ is a `real' unit ({\it ie}, $b=\sqrt{a}\not=\sqrt{-a}$).}.
Also, by comparing the multiplication tables (\ref{eq13}) and
(\ref{eq15}) for $\A$ and $\quat$ respectively, one immediately
realizes that the former is a sort of {\em deformation} of the
latter\footnote{With most notable `deformation features' of the
generators of $\A$ relative to those of $\quat$ being $c$'s
squaring to $a$ unlike $j$'s squaring to $-1$ mentioned in the
last footnote, and $a$'s role only as a right-identity unlike
$1$'s role in $\quat$ as being both a right and a left-identity.
In fact, from the diagonals of their respective multiplication
tables (\ref{eq15}) and (\ref{eq13}), one could say that the unit
quaternions in $\mathbf{u}$ naturally support a metric of {\em
Lorentzian signature} $\mathrm{diag}(1,-1,-1,-1)$ (absolute trace
2) \cite{lambek, trif}, while the units of $\A$ in $\mathbf{g}$
support a metric of {\em Kleinian signature}
$\mathrm{diag}(1,-1,1,-1)$ (traceless). Otherwise, see
correspondence (\ref{eq14}) in section \ref{sec3}.}.

\item (d)  Like $\cayl$, $\A$ is not associative.

\item (e) Furthermore, the novel multiplicatively ordered
$\Z_{2}$-graded associative structure of $\A$ recalls a bit the
linearly ordered $\R$. Could a structure like that be used to
define somehow a $\D$-like ring, thus extend Hurwitz's theorem in
\cite{hur} to $16$ dimensions?\footnote{For example, since the
extension from $\R$ to $\com$, to $\quat$, and finally, to
$\cayl$, involves a complexification-like process ({\it ie}, one
adjoins an imaginary unit to the existing generators and demands
algebraic closure---thus, in effect, one doubles the dimension),
further extension of $\cayl$ to $\D$ could involve an `algebraic
degree of freedom' coming from $\Z_{2}$-grading ({\it ie}, somehow
assume that the usual octonion units are {\em even} and that the
other eight needed to comprise the decahexanions are {\em odd}, so
that some kind of $\Z_{2}$-graded associative multiplication {\it
\`a la} $\A$, not covered by Hurwitz \cite{hur}, could be evoked;
for instance, one could assume that the $\jmath$ ($\jmath^{2}=-1$)
adjoined to the eight `even' octonion units $\{
1,\imath_{1},\cdots ,\imath_{7}\}$ is odd which, by the displayed
expression at the top of page 6, would make the other seven
resulting decahexanion units $\{ \imath_{1}\jmath ,\cdots
,\imath_{7}\jmath\}$ odd as well). However, at this stage this is
purely `heuristic speculation'.}.

\end{itemize}

\vskip 0.1in

$\mathbf{(2)}$ We stressed above the close similarities between
$\A$ and $\quat$. Now we would like to gain some more insight into
the novel non-associativity of $\A$ by comparing it with the
associative quaternions. As a bonus from such a comparison, we
will also comment briefly on a possible representation of $\A$.

So, we may recall from \cite{lambek} the real 4-dimensional left
($L$) and right ($R$) matrix `self-representations'\footnote{The
epithet `self' refers to the representation of $\quat$ (by real
matrices) induced by the quaternions' own algebraic product.} of
quaternions over $\R$

\begin{equation}\label{eq27}
{\rm Left:}~ab=c\mapto L(a)[b]=[c]~~{\rm and}~~{\rm
Right:}~bc=d\mapto R(c)[b]=[d],
\end{equation}

\noindent where $[b]$ is a column vector in $\R^{4}$\footnote{That
is, in the expansion of the real quaternion $b$ in the standard
unit quaternion basis $\mathbf{u}$:
$b=b_{0}1+b_{1}i+b_{2}j+b_{3}k$, the entries of the $4$-vector
$[b]$ are the real numbers $b_{\mu}$.}, while both $L(a)$ and
$R(c)$ are $4\times 4$ real matrices\footnote{It is easy to check
that the maps $L$ and $R$ are homomorphisms of $\quat$ ({\it ie},
representations of $\quat$).}. The crucial point is that, because
$\quat$ is associative,

\begin{equation}\label{eq28}
(ab)c=a(bc)\Rightarrow R(c)L(a)[b]=L(a)R(c)[b]\Leftrightarrow
[L(\quat),R(\quat)]=0,
\end{equation}

\noindent and similarly, for a purely antiassociative algebra like
$\jor$ before, it follows that

\begin{equation}\label{eq29}
\{ L(\jor), R(\jor)\}=0.
\end{equation}

\noindent We may summarize (\ref{eq28}) and (\ref{eq29}) to the
following:

\begin{quotation}

\noindent The left and right self-representations of an {\em
associative} algebra {\em commute}, while those of an {\em
antiassociative} algebra {\em anticommute}.

\end{quotation}

\noindent It follows that the self-representations of $\A$, which
is neither purely associative nor purely associative (but somewhat
in between the two), will neither commute nor anticommute with
each other. As a matter of fact, since $\A$ is multiplicatively
ordered $frtl$, only its left self-representation would be
relevant (if it actually existed\footnote{This author has not been
able to construct yet a matrix representation of $\A$ based on its
ordered $\Z_{2}$-graded associative product. Of course, like with
all the usual Lie algebraic varieties and supervarieties, we could
alternatively look directly into a possible representation of the
non-associative (under the Lie bracket $<.,.>$ now!) $\A$ by a
(possibly graded) Lie algebra $End(V)$ of endomorphisms of a
suitable vector space $V$. However, this alternative has not been
seriously explored yet.}).

\vskip 0.1in

$\mathbf{(3)}$ The abstraction of $\A$ to a general mixed
$\de$-J-L algebra $\JL$ is straightforward:

\begin{quotation}

\noindent A finite dimensional $\Z_{2}$-graded vector space $\JL$
over a field $K$ of characteristic not $2$, together with an
ordered $\Z_{2}$-graded associative multiplication between its
elements and a $\Z_{2}$-graded Lie-like bracket $<.,.>$ satisfying
(\ref{eq21})--(\ref{eq25}), is called an {\em abstract mixed
$\de$-Jordan-Lie superalgebra}.

\end{quotation}

$\mathbf{(4)}$ We conclude the present paper by allowing ourselves
some latitude so as to discuss briefly a possible physical
application and concomitant interpretation of $\A$.

\vskip 0.07in

$\A$ was originally conceived in \cite{rap1}, but not in the
rather sophisticated $\de$-J-L guise presented above. The basic
intuition in \cite{rap1} was to give a simple `{\em generative
grammar}'-like theoretical scenario for the creation of spacetime
from a finite number of quanta (generators) which were supposed to
inhabit the quantum spacetime substratum commonly known as the
vacuum \cite{df91}. Thus, it was envisaged that a spacetime-like
structure could arise from the algebraic combinations of a finite
number of quanta, as it were, a combinatory-algebraic process
modelling the {\it aufbau} of spacetime from quantum spacetime
numbers filling the vacuum\footnote{Thus, $\A$ could be coined
`quantum spacetime arithmetic' and the imagined process of
building spacetime from such abstract numbers is akin, at least in
spirit, to how relativistic spacetime was assembled from abstract
digits and a suitable code or `algorithm' for them in
\cite{df69}.}. Furthermore, by the very alphabetic character of
$\A$ and its alphabetically ordered algebraic structure, this
syntactic {\em lexicographic} process representing the building of
spacetime was envisaged to encode the germs of the primordial
`{\em quantum arrow of time}' in the sense that a primitive
`temporal directedness' is already built into the algebraic
structure of those quantum spacetime numbers---a basic order or
`{\it taxis}' inherent in the very rules for the algebraic
combinations of the generators of $\A$, as we saw before. In view
of the intimate structural similarities between $\A$ and the
quaternion division algebra $\quat$ mentioned above, and since the
latter are so closely tied to the structure of relativistic
spacetime and the best unification between quantum mechanics and
relativity that has been achieved so far, namely, the Dirac
equation \cite{lambek}\footnote{For example, in \cite{lambek}
Minkowski vectors are represented by hermitian biquaternions,
Lorentz transformations by unimodular complex quaternions
(essentially, the biquaternion analogues of the elements of
$SL(2,\com)$---the double covering of the Lorentz group), the
$3$-generators $\sigma_{i}$ of the Pauli spin Lie algebra $su(2)$
are just the three mutually anticommuting `imaginary' quaternions
multiplied by the complex number $i$ in front ($i^{2}=-1$), and,
most importantly, the Dirac equation can be derived very simply
and entirely algebraically from $\quat$ over $\com$ ({\it ie},
from biquaternions). Also, as noted in footnote 40, the Lorentzian
signature (and even the dimensionality!) of Minkowski spacetime is
effectively encoded in (the diagonal of) the multiplication table
(\ref{eq15}) of the unit quaternions in $\mathbf{u}$
\cite{trif}.}, we can imagine that $\A$ could be somehow used in
the future to represent algebraically a `time-directed' sort of
Minkowski spacetime and a time-asymmetric version of the Dirac
equation that would appear to be supported rather naturally by the
former. However, the quest in this direction is far from its
completion.

\vskip 0.05in

We would like to close the present paper in the spirit of the last
paragraph with a suitable quote from the end of \cite{kauff}:

\vskip 0.1in

\centerline{``In the beginning was the word.}

\centerline{The word became self-referential/periodic.}

\centerline{In the sorting of its lexicographic orders,}

\centerline{The word became topology, geometry and}

\centerline{The dynamics of forms;}

\centerline{Thus were chaos and order}

\centerline{Brought forth together}

\centerline{From the void.''}

\centerline{(from {\bf CODA})}

\section*{Acknowledgments}

Some early remarks by Jim Lambek about $\quat$ in \cite{lambek}
and in subsequent private correspondence helped this author
clarify by analogy some crucial structural issues about $\A$. This
author acknowledges the generous material support by the European
Commission in the form of a Marie Curie Individual Postdoctoral
Research Fellowship held at Imperial College, London.


\begin{thebibliography}{99}

\bibitem{df69} Finkelstein, D., {\itshape Space-Time Code}, Physical
Review, {\bf 184}, 1261 (1969).

\bibitem{df91} Finkelstein, D., {\itshape Theory of Vacuum}, in
{\itshape The Philosophy of Vacuum}, Eds. Brown, H. and Saunders,
S., Clarendon Press, Oxford (1991).

\bibitem{freund} Freund, P. G. O., {\itshape Introduction to Supersymmetry},
Cambridge University Press, Cambridge (1989).

\bibitem{hur} Hurwitz, A., {\itshape \"{U}ber die Composition der quadratischen
Formen von Beliebig vielen Variablen}, Nachrichten von der
k\"{o}niglichen Gersellschaft der Wissenshaften in G\"ottingen
(1898).

\bibitem{kauff} Kauffman, L. H., {\itshape Knots and Physics},
Series on Knots and Everything, volume 1, World Scientific,
Singapore (1991).

\bibitem{lambek} Lambek, J., {\itshape If Hamilton had prevailed:
quaternions in physics}, Mathematical Intelligencer, {\bf 17}, 7
(1995).

\bibitem{okam} Okubo, S. and Kamiya, N., {\itshape Jordan-Lie Super Algebra and
Jordan-Lie Triple System}, Journal of Algebra, {\bf 198}, 388
(1997).

\bibitem{pauli} Pauli, W., {\itshape On the Connection between
Spin and Statistics}, Physical Review, {\bf 58}, 716 (1940).

\bibitem{rap1} Raptis, I., {\itshape Axiomatic Quantum Timespace Structure:
A Preamble to the Quantum Topos Conception of the Vacuum}, Ph.D.
Thesis, University of Newcastle upon Tyne, UK (1998).

\bibitem{trif} Trifonov, V., {\itshape A Linear Solution of the
Four-Dimensionality Problem}, Europhysics Letters, {\bf 32}, 621
(1995).

\bibitem{www} Wick, G. C., Wightman, A. S. and Wigner, E. P.,
{\itshape The intrinsic parity of elementary particles}, Physical
Review, {\bf 88}, 101 (1952).

\end{thebibliography}
\end{document}